\documentclass[a4paper]{article}

\usepackage{graphicx}

\usepackage{soul}
\usepackage{graphicx}
\usepackage{epstopdf, epsfig}
\usepackage{amsmath}
\usepackage[table]{xcolor}
\usepackage{subcaption}
\usepackage{tabularx}
\usepackage{soul}
\usepackage{makecell}
\usepackage{multirow}

\usepackage{breqn}

\newcommand{\ah}[1]{{\color{black}#1}} 
\newcommand{\rev}[1]{{\color{black}#1}} 
\newcommand{\soptitle}{Physics-Informed Scaling Laws for the Performance of Pitching Foils in Schooling Configurations}

\colorlet{Purple}{blue!40!red}

\begin{document}


\begin{center}
\Large \bf{\soptitle}
\vspace{0.1in}
\end{center}

\begin{center}
{Ahmet Gungor$^1$, Muhammad Saif Ullah Khalid$^{1,2}$, and Arman Hemmati$^{1\star}$}\\
\vspace{0.1in}
\end{center}
\begin{center}
$^1$Department of Mechanical Engineering, University of Alberta, Edmonton, Alberta T6G 2R3, Canada\\
$^2$Department of Mechanical and Mechatronics Engineering, Lakehead University, Thunder Bay, ON P7B 5E1, Canada\\
\vspace{0.05in}
$^\star$\small{Corresponding Author, Email: arman.hemmati@ualberta.ca}
\end{center}

\begin{abstract}
This study introduces novel physics-based scaling laws to estimate the propulsive performance of synchronously pitching foils in various schooling configurations at $Re=4000$. These relations are derived from  quasi-steady lift-based and added mass forces. Hydrodynamic interactions among the schooling foils are considered through vortex-induced velocities imposed on \ah{them, constituting the ground effect.} Generalized scaling equations are formulated for cycle-averaged coefficients of thrust and power. These equations encompass both the pure-pitching and induced velocity terms, capturing their combined effects. The equations are compared to computational results obtained from two-foils systems, exhibiting \ah{foil arrangements over a wide range of parameter space,} including Strouhal number ($0.15 \leq St \leq 0.4$), pitching amplitude ($5^\circ \leq \theta_0 \leq 14^\circ$), and phase difference ($0^\circ \leq \phi \leq 180^\circ$). The individual contributions of pure-pitching and induced velocity terms to propulsive performance elucidate that solely relying on the pure-pitching terms leads to inadequate estimation, emphasizing the significance of the induced velocity terms. Validity of the approach is further assessed \ah{by testing it with} a three-foil configuration, which displays a collapse. This indicates that the scaling laws are not only applicable to two-foils systems but also extend their effectiveness to multi-foil arrangements.

\end{abstract}

\section{Introduction}

Employing bio-inspired approaches can play a vital role in designing highly efficient underwater swimming robots. Physical mechanisms that govern various important swimming strategies of natural aquatic species provide crucial knowledge for improving relevant engineering designs. Oscillating foils are often used as a model to realize the motion of the backbone of fish and their caudal fins \cite{triantafyllou-efficienct-swimming-machine-1995, triantafyllou-fishlike-swimming-2000}. Wake dynamics and propulsive performance of solitary oscillating foils were extensively studied recently \cite{floryan-efficient-cruising-2018, anderson1998oscillating, verma_vortex_evolution}.  

 Fish schooling is a ubiquitous phenomenon that natural species exhibit for different reasons, spanning from social benefits to hydrodynamic advantages \cite{shaw-fish-school-1962}. There exists overwhelming evidence that fish enhances their swimming performance through schooling. Ashraf et al.\cite{ashraf-syncronization-2016, ashraf-simple-phalanx-2017} demonstrated that two red nose tetra fish positioned themselves nearly side-by-side and synchronized the beating frequencies of their tail fins, when forced to swim faster in a tank. Similarly, it has been demonstrated that oscillating multiple foils achieved an improved propulsive performance by leveraging side-by-side\cite{gungor-asymmetry, gungor-fish-swimming, dewey-propulsive-performance-2014}, in-line \cite{boschitsch-in-line-tandem-2014, pourfarzan_in-line_performance}, or staggered \cite{huarte-staggered-foils-2018} configurations, depending on the spacing between the foils. For example, our study of the propulsive performance of pitching foils in side-by-side configuration\cite{gungor_phase_difference_2022} revealed that in-phase and out-of-phase pitching motion of the foils improved their efficiency compared to a single pitching foil. However, intermediate phase differences appeared to reduce their efficiency. 

With the growing interest of the scientific community in this area, analytical models were developed to quantify and analyze the metrics of propulsive performance of oscillating foils and plates. Using potential flow theory and the Kutta condition, Theodorsen\cite{theodorsen-aeorodynamic-1935} derived linearized models for fluidic forces around an oscillating airfoil. Garrick\cite{garrick_scaling} examined oscillating foils and foil-aileron combinations, based on which they constructed formulations for the propelling forces generated by the system and its corresponding power input and output. Floryan et al.\cite{floryan-scaling-propulsive-performance-2017} derived scaling relations for cycle-averaged thrust, power, and efficiency of an isolated foil, undergoing heaving and pitching motion. It was further demonstrated that these mathematical models produced results that were consistent with the biological data for aquatic animals. Moored and Quinn\cite{moored_inviscid_scaling_2019} argued that Garrick's theory was able to capture mean thrust forces of the foils, whereas it poorly estimated the mean power. They provided a corrected power scaling for self-propelling pitching foils, which incorporated the effects of added-mass forces and large-amplitude shear layer separated from the trailing edge. They suggested that separating shear layer contributed to the scaling of power through its circulation and vortex proximity. Ayancik et al.\cite{ayancik-3D_scaling-2019, ayancik-fluke-3Dscaling-2020} focused on pitching propulsors of finite spans and modified the formulations provided earlier by Moored and Quinn\cite{moored_inviscid_scaling_2019}. The new scaling \ah{formulation} included modified terms that capture the effects of three-dimensional flows on  performance parameters \ah{based on the foil aspect ratio}. Ayancik et al.\cite{ayancik-3D_scaling-2019, ayancik-fluke-3Dscaling-2020} provided insights on the physical mechanisms responsible for producing forces by catacean flukes. This allowed determining an optimal non-dimensional heaving amplitude that maximizes the propulsive efficiency of oscillating foils. Similarly, Akoz and Moored \cite{akoz_intermittent_2018} altered the scaling relations for intermittently pitching foils and explained the reasons for hydrodynamic benefits of intermittent swimming using those equations. Further expanding this work, Floryan et al.\cite{floryan-scaling-propulsive-performance-2017} and Van Buren et al.\cite{buren-scaling-2019} presented scaling laws for simultaneously heaving and pitching foils. In  another study, Van Buren et al.\cite{buren_flow_speed_2018} performed scaling-based analysis \ah{on the same kinematics}, which revealed that streamwise speed of the flow had little or no impact on the performance of oscillating foils. This was related to the lateral velocity of the foils, which dominated the forces. Thus, they argued that performance-related conclusions drawn from tethered foils should be applicable to free-swimming conditions. 

Previous work on developing scaling relations for performance metrics of oscillating foils was primarily focused on single isolated foils; however, there exist limited studies about constructing such formulations for multiple foils. Recently, Quinn et al. \cite{quinn-solid-boundary-2014} investigated the propulsion of a pitching foil in ground effect, which effectively makes it a system composed of two foils in a side-by-side arrangement with out-of-phase synchronization. They suggested an empirical power-law scaling for both thrust and power. Mivechi et al. \cite{mivechi_scaling_ground_2021} addressed the same problem using physics-based modifications to the added mass and circulatory forces to include the ground effect by incorporating the edge vortex and its image. \ah{In another study} Simsek et al. \cite{simsek-scaling-in-line-2020} presented scaling relations for two in-phase pitching foils in an in-line configuration. More recently, Gungor and Hemmati \cite{gungor-scaling} developed scaling equations for propulsive performance of pitching foils in side-by-side configurations, utilizing empirical expressions for spacing between the foils and their phase difference. However, limitations of these very recent research efforts in terms of their slender parametric space or more reliance on empirical formulations are clear. In order to design bio-inspired swimming robots, it is critical to construct physics-informed scaling laws applicable to a broad range of design and schooling parameters, including relative positions of oscillating foils and their phase differences. This primarily lays the foundation and motivation for the present study to develop scaling relations for a wide parametric space, including oscillation amplitude, oscillation frequency, separation distance, and phase difference. It is important to mention that developing universal scaling laws applicable to a variety of fish schooling configurations needs incorporation of flow conditions (Reynolds numbers), geometric characteristics (separation distance between the members of a school), kinematic parameters (oscillation frequencies and amplitudes, and phase difference), and physiological features of different aquatic species. In this context, our newly proposed scaling relations for the propulsive metrics of individual members cover a wide, though not full, spectrum of the governing conditions in fish schools. Our manuscript is structured as follows: Section \ref{ref_methodology} provides an overview of the problem definition, introduces the parametric space, and outlines the computational methodology employed. In Section \ref{scaling_laws}, we delve into the derivation of the novel scaling laws for staggered foils, highlighting their formulation and underlying principles. The corresponding results obtained from our analysis are presented in Section \ref{results}. Finally, we draw conclusions from the study and summarize our findings in Section \ref{conclusion}.

\section{Methodology}
\label{ref_methodology}

Propulsive performance of pitching foils were evaluated by directly solving the Navier-Stokes equations in OpenFOAM. Rigid teardrop foils with chord length of $1c$ and semi-circular leading edge with a radius of $0.05c$ were placed in a uniform flow. Sinusoidal pitching motion was imposed on the foils, which is described as: 
\begin{equation}
\theta(t)=\theta_0\sin(2\pi ft -\phi),
\label{eq_theta}
\end{equation}
Here, $\theta_0$ is the amplitude of pitching, $t$ is time, $f$ is the pitching frequency and $\phi$ is the phase difference between the pitching motion and a sine wave. Non-dimensional parameters used in the analysis are given below:
\begin{equation}
 St=\frac{f A}{U_{\infty}}, \;\;\;\;\;  f^*=\frac{fc}{U_{\infty}}, \;\;\;\;\;   A^*=\frac{A}{c}, \;\;\;\;\;  Re=\frac{U_{\infty}c}{\nu}, 
\end{equation}
\noindent where $St$ is Strouhal number, $U_{\infty}$ is the free-stream flow velocity, $A$ is the two-sides amplitude of oscillation, $f^*$ is the reduced frequency, $A^*$ is the non-dimensional pitching amplitude, $Re$ is Reynolds number, and $\nu$ is kinematic viscosity of the fluid.

\begin{figure}
	\centering
	\includegraphics[width=0.8\textwidth]{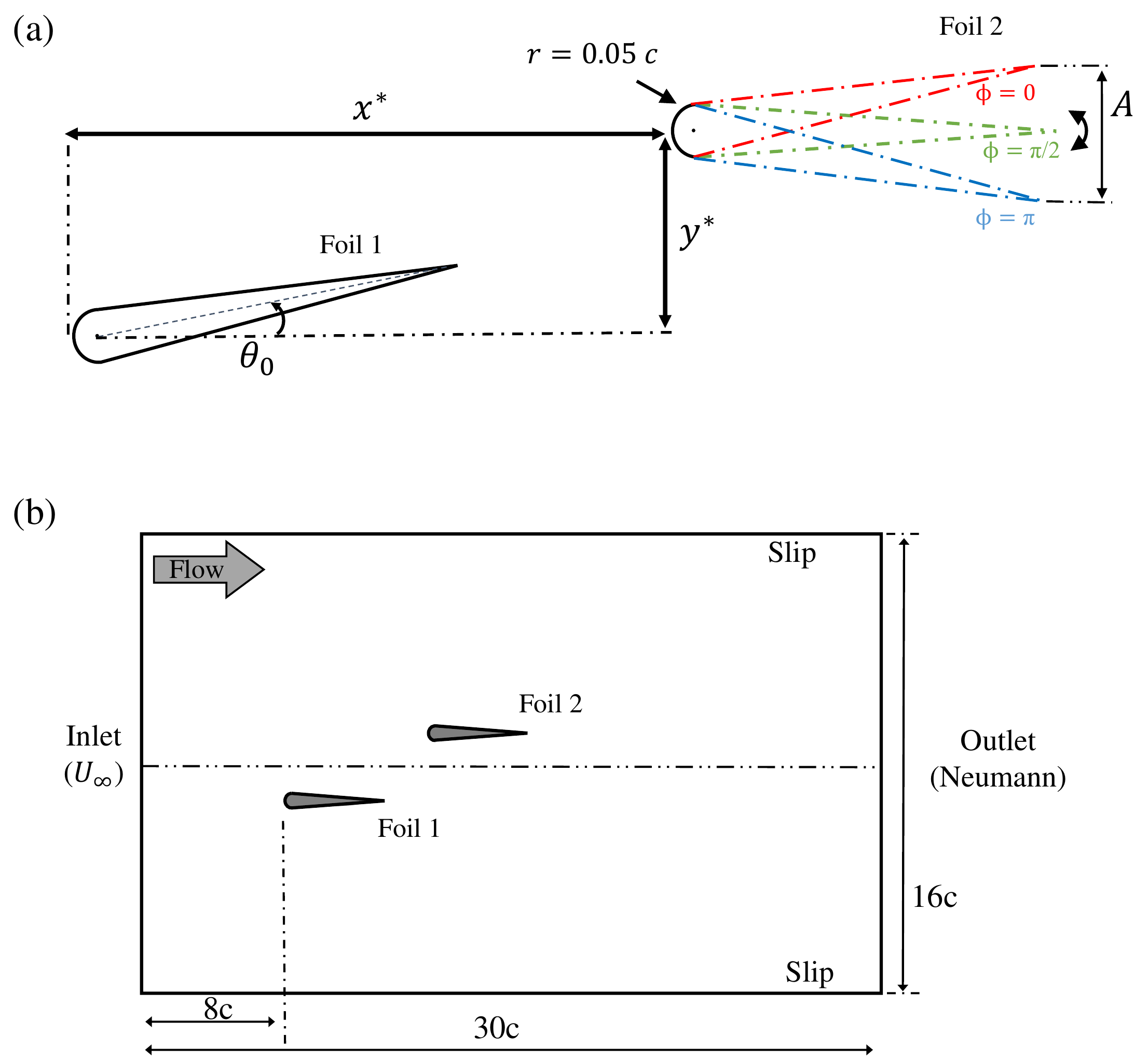}
	\caption{Demonstration of (a) the pitching motion of two foils in staggered configuration (b) two-dimensional computational domain with boundary conditions (not to scale)}
	\label{fig_domain}
\end{figure}

The computational domain was rectangular extending $30c$ in the streamwise direction ($x-$) and $16c$ in the cross-flow direction ($y-$). A uniform velocity boundary condition ($u=U_\infty,v=w=0)$ was applied at the inlet, which was located 8c away from the leading edge. The foils surfaces were subject to a no-slip wall boundary condition, while the upper and lower boundaries of the domain utilized a slip boundary condition as described in Hemmati et al. \cite{hemmati-trailing-edge-shape-2019}. Neumann boundary condition for both pressure and velocity were imposed at the outlet

To accurately model the flow around foils, a non-homogeneous spatial grid with at least $~7.13\times10^5$ hexahedral elements was utilized. A finer grid was placed around the foils, while a coarser mesh was located close to the boundaries. A total of 600 nodes were placed on the foils, which is a finer grid resolution than what was used by Senturk and Smits \cite{senturk-reynolds-scaling-2019}. The grid size expands gradually towards the boundaries, while ensuring that the expansion ratio does not exceed 1.03 throughout the computational domain. The incompressible, transient flow solver for moving mesh cases, pimpleDyMFoam, was used. It employs PIMPLE algorithm, a hybrid of PISO (Pressure-Implicit with Splitting of Operators) and SIMPLE (Semi-Implicit Method for Pressure Linked Equations). To ensure numerical stability, the time-step size was selected to meet the CFL-condition, which remained below 0.80 in this study. Second-order implicit time-marching schemes were used to discretize temporal terms, and convective and diffusive terms were approximated using second-order accurate schemes. The mesh morphing method was used to impose the pitching motion, and as a result, the solver deforms the grid around the two pitching foils at every time-step while maintaining grid quality.

\begin{table*}
	\caption{\label{table_parameter_space} Parametric space of the study.}
	\centering
	\renewcommand{\arraystretch}{1.5}
	\setlength{\tabcolsep}{8.2pt}
	\begin{tabular}{ccccc}
            \hline
            Parameter &  Range & Increments \\
		\hline
		$x^*$ &  0c $-$3c & 0.5c\\
		$y^*$  & 0.5c$-$2c & 0.5c\\
        $St$ &  0.15$-$0.4& 0.05	\\
         $Re$ &  4000 & $-$\\
		$\phi$	& $0^\circ$ $-$ $180^\circ$	& $30^\circ$ \\
		$\theta_0$	& $5^\circ$ $-$ $14^\circ$ & $3^\circ$\\

		\hline
	\end{tabular}
\end{table*}
 
A wide range of spatial configurations were simulated in this study. Although we mostly conducted the numerical simulations for two-foils systems, a configuration with three pitching foils were also examined to demonstrate the extensiveness of the scaling approach. The horizontal and vertical separation distances between the foils, denoted by $x^*$ and $y^*$, respectively, were varied between $0.5c-2c$ and $0.5c-3c$, respectively, with increments of $0.05c$. The amplitude-based Strouhal number of the flow was set to $St=0.15-0.4$, which encompasses the range in which fish naturally swim \cite{godoy-transitions-wake-2008, triantafyllou-optimal-thrust-1993}. Additionally, the phase difference between the foils was varied from $\phi=0$ (in-phase motion) to $\phi=\pi$ (out-of-phase motion), with increments of $\pi/6$. The pitching amplitude was also varied between $5^\circ \leq \theta \leq 14^\circ$ with increments of $3^\circ$. A summary of the parameter space used in this study is provided in Table \ref{table_parameter_space}.

The performance of oscillating foils is commonly evaluated using coefficients of thrust ($C_T$) and power ($C_P$). These coefficients are defined by 
\begin{equation}
{C_T}=\frac{{F_x}}{{\textstyle \frac{1}{2}} \rho U_\infty^2 sc},\label{eq_ct}
\end{equation}	
\begin{equation}
{C_P}=\frac{{M_z}\dot{\theta}}{{\textstyle {1 \over 2}} \rho U_\infty^3 sc},\label{eq_cp}
\end{equation}
where $F_x$ is the streamwise force on the foil, $M_z$ is the moment about the z-axis experienced by a foil, $\rho$ is the fluid density, and $s$ is the span of the foil. To obtain the mean coefficients of thrust and power, $\overline{C_T}$ and $\overline{C_P}$ were averaged over $10$ pitching cycles after achieving statistical convergence in the simulations. To quantify statistical convergence, we calculated the percent change between consecutive oscillation cycles and ensured that it remained below $1\%$ for the last $10$ cycles. Then, the propulsive efficiency can be calculated as the ratio of $\overline{C_T}$ to $\overline{C_P}$, given as:
\begin{equation}
{\eta}=\frac{\overline{C_T}}{\overline{C_P}}.\label{eq_eff}
\end{equation}
To ensure temporal accuracy in the simulations, each oscillation cycle consisted of more than $3500$ time-steps. A comprehensive sensitivity analysis of grid, time-step, and domain sizes, along with a validation study of our computational methodology, can be found in Gungor et al. (2021) \cite{gungor-scaling}.

\section{Scaling laws}
\label{scaling_laws}
In this section, we elucidate the construction of a new physics-based scaling formulation for propulsion metrics, thrust and power, for synchronously pitching foils. We consider a wide range of schooling arrangements and design parameters. The term 'synchronous' implies that the foils in the system utilize the same pitching frequency. Pitching frequencies of the foils are kept the same, following the natural swimming habits of various fish species swimming in schooling configurations. Different biological studies \cite{herskin1998energy, svendsen2003intra} revealed that the following members of fish schools could reduce their energy consumption by beating their tails with lower frequencies than those of their leaders. However, the difference in frequencies was found to be less than $15\%$ \cite{herskin1998energy, svendsen2003intra}. From Khalid et al. \cite{khalid2016hydrodynamics} and Newbolt et al. \cite{newbolt2019flow}, it is evident that there exists a very small range of governing kinematic parameters that allow for hydrodynamic advantages for follower fish, which led to the formation of stable configurations in fish schools. Additionally, oscillation frequencies of schooling red nose tetra fish are measured to be in close proximity in both two-fish and three-fish schools \cite{ashraf-syncronization-2016}. These observations serve as a justification for considering synchronized kinematics for the members of fish schools in this study. 

\begin{figure}
	\centering
	\includegraphics[width=1\textwidth]{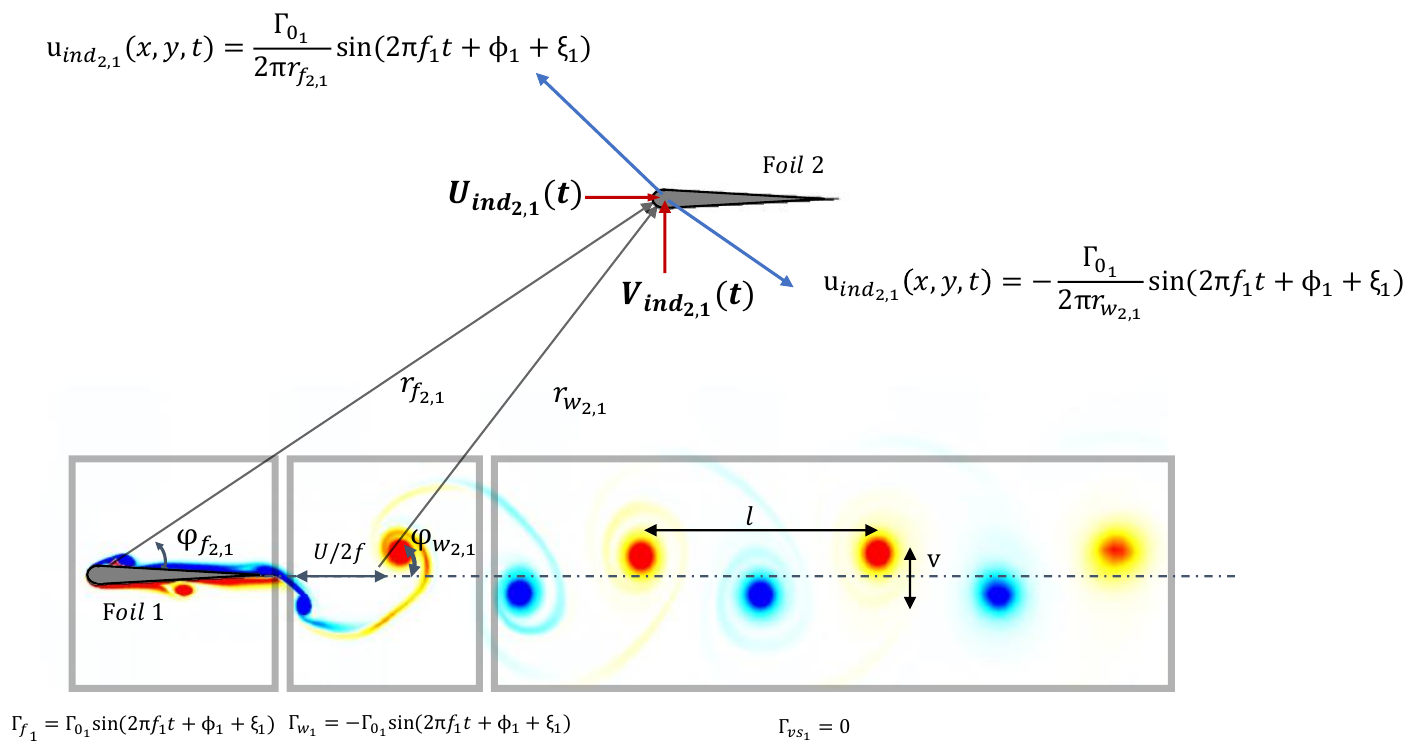}
	\caption{Induced velocities on the follower foil due to the circulation around and in the wake of the leader foil, together with the geometric quantities used in the analysis. Here, $\Gamma_f$, $\Gamma_0$, $\Gamma_{w}$, and $\Gamma_{vs}$ represent the circulation around the foil, the amplitude of circulation, the circulation in the wake, and the circulation in the vortex street, respectively.}
	\label{fig_circulation}
\end{figure}

\subsection{Scaling approach}
We employ an approach similar to Floryan et al.\cite{floryan-scaling-propulsive-performance-2017} and Van Buren et al. \cite{buren-scaling-2019}, who established scaling relations for isolated foils based on lift-based forces \cite{theodorsen-aeorodynamic-1935}, added mass forces \cite{sedov-two-dimensional-1965}, and contributions from the drag due to fluid flow over the oscillating bodies. A novel element of our present work originates from our analysis considering the impact on a foil through the vortex-induced velocity imposed by the other foils in the schooling configuration and their wakes. We derive the scaling equations not for each foil individually but in a generalized manner. To this end, we utilize the subscript $i$ to represent the foil number, while double subscripts $i$ and $j$ are employed to describe the interaction between the foils and its direction. For instance, $\overline{C_{T_{i}}}$ corresponds to the cycle-averaged thrust of Foil $i$, whereas $V_{ind_{i,j}}$ defines the velocity on Foil $i$ induced by Foil $j$. It is important to mention that we employ small-angle approximation throughout the derivation process for our scaling formulations. 

Now, we begin with the description of lift-based forces and their contribution to the scaling relations. Pitching foils produce unsteady lift that follows their sinusoidal motion, $\theta_{i}(t)=\theta_{0}\sin(2\pi f_i t -\phi_i)$, but with a phase lag $\xi_i$ that depends on the reduced frequency $f^*$. This phase lag indicates the lag \ah{of lift data compared to} the pitching motion of the foil. This is a consequence of the time taken during the development of the vortex shedding process. The relationship between the phase lag and the reduced frequency can be estimated by Theodorsen's reduced order model \cite{theodorsen-aeorodynamic-1935}, which provides an accurate approximation except for very small reduced frequencies \cite{chiereghin_lift_phase_2019}. Hence, unsteady lift force on a foil is expressed as $L_{uns_{i}}=L_{0_{i}}  \sin(2 \pi f_i t-\xi_i)$, where $L_0$ is the amplitude of lift. Since the lift force is proportional to the circulation around the foil through the Kutta-Joukowski theorem (\textit{i.e.,} $\Gamma_i =L_{uns_{i}}/ \rho s U_{\infty}$), it induces a momentum (velocity) on the other foil in its vicinity. Employing the Kutta condition, it can be assumed that circulation with the same strength but opposite sign is developed in the wake (please see Fig.~\ref{fig_circulation}). This circulation is centered in the immediate vicinity of the trailing edge since the circulations of oppositely sign vortices in the rest of the wake cancel each other. The underlying reason for this cancellation is their equal strength that is opposite in sign at moderate $St$ \cite{godoy-model-symmetry-breaking-2009}. We estimate the center of wake circulation to be located $U/2f$ from the trailing edge, as this is the lateral distance between two consecutively shed vortices, also quantifying the wavelength of the near-wake. While it is true that the vertical location of the circulation may not align with the wake centerline for cases where the wake is deflected \cite{gungor_wake_merging}, we expect its impact on the induced velocity to be minimal. It is because the shift in the vertical location of vortices is considerably smaller compared to the distance between the foils for any possible wake topology \cite{gungor_wake_merging, gungor-asymmetry, gungor_phase_difference_2022}. Circulation around a foil and in its wake is counter-productive since they possess opposite-signs. Their combined effect induces a velocity in both vertical and horizontal directions. Thus, these induced velocities allow us to consider swimmer's performance as a single foil, which undergoes combined heaving and pitching motion. This argument is based on the induced velocity in vertical direction that may be analogous to heave velocity of the simultaneously heaving and pitching foil. Hence, we can compute the induced velocities in the $x-$ and $y-$ directions for a foil using the Biot-Savart rule as follows:
\begin{align}
V_{ind_{i}} & = \sum\limits_{\substack{j=1 \\ j\neq i}}^n \frac{\Gamma_j}{2 \pi} \left [\frac{\cos\psi_{f_{i,j}}}{|r_{f_{i,j}}|}-\frac{\cos\psi_{w_{i,j}}}{|r_{w_{i,j}}|} \right ] \nonumber\\
& = \sum\limits_{\substack{j=1 \\ j\neq i}}^n \frac{L_0 \sin(2 \pi f t -\xi_j -\phi_j)}{2 \pi \rho s U_{\infty}} \left [\frac{\cos\psi_{f_{i,j}}}{|r_{f_{i,j}}|}-\frac{\cos\psi_{w_{i,j}}}{|r_{w_{i,j}}|} \right ]\label{eq_v_ind}\\
U_{ind_{i}} & =  \sum\limits_{\substack{j=1 \\ j\neq i}}^n \frac{\Gamma_j}{2 \pi} \left [\frac{\sin\psi_{f_{i,j}}}{|r_{f_{i,j}}|}-\frac{\sin\psi_{w_{i,j}}}{|r_{w_{i,j}}|} \right ] +U^*_{i,j} \nonumber \\
& = \sum\limits_{\substack{j=1 \\ j\neq i}}^n \frac{L_0 sin(2 \pi f t -\xi_j -\phi_j)}{2 \pi \rho s U_{\infty}} \left [\frac{sin\psi_{f_{i,j}}}{|r_{f_{i,j}}|}-\frac{sin\psi_{w_{i,j}}} {|r_{w_{i,j}}|} \right ] +U^*_{i,j}
\label{eq_u_ind}
\end{align}
\noindent where n denotes the total number of foils in the school, $\Gamma_j$ is the quasi-steady sinusoidal circulation generated by the foils, $r_{f_{i,j}}$ is the radial distance vector \ah{(from the center of the circulation around Foil $j$ to the leading edge of Foil $i$), $r_{w_{i,j}}$ is the radial distance vector (from the center of the wake circulation of Foil $j$ to the leading edge of Foil $i$), $\psi_{f_{i,j}}$ denotes the angle from the center-line of Foil $j$ to $r_{f_{i,j}}$, $\psi_{w_{i,j}}$ denotes the angle from the center-line of Foil $j$ to $r_{w_{i,j}}$,} and $U^*_{i,j}$ shows the time-independent velocity contribution on Foil $i$ due to the vortex street of Foil $j$ that is further explained in the following paragraph.

\begin{figure}
	\centering
	\includegraphics[width=0.7\textwidth]{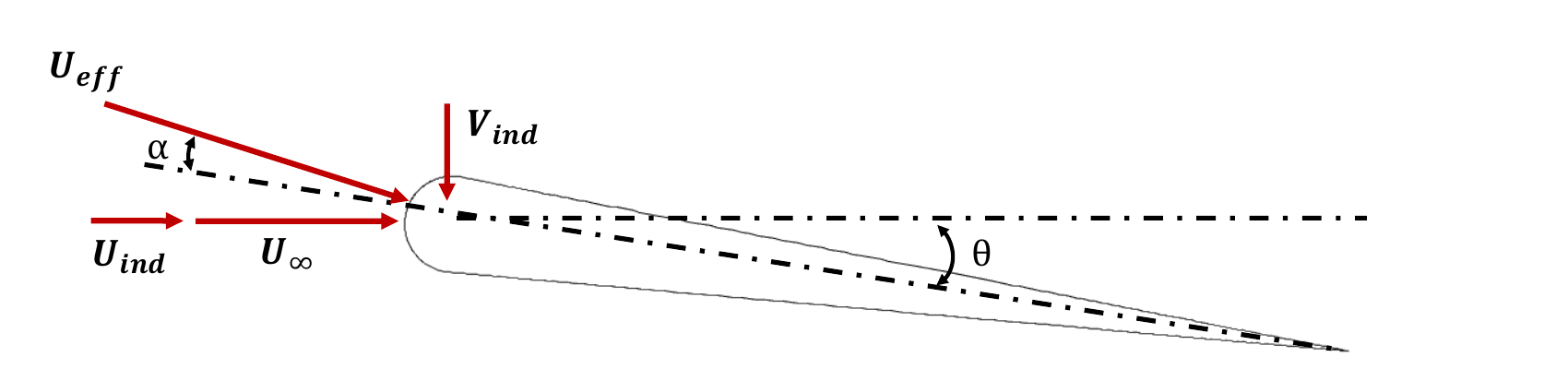}
	\caption{Velocity components arise on the pitching foils in staggered configurations.}
	\label{fig_velocity}
\end{figure}

In addition to the velocities induced by circulation around the foil and its counterpart in the immediate wake, the rest of the wake also contributes to the induced velocities. It forms a vortex street with apparently net zero circulation as previously explained. However, the vortices in the vortex street have varying vertical distances from the foil; therefore, their total contribution to the induces velocities can be non-zero. Schaefer and Eskinazi \cite{schaefer_vortex_street_1959} and Griffin and Ramberg \cite{griffin_vortex_street_1974} provided potential flow solutions for velocities at a point induced by an infinite vortex street generated by tripping and vibrating cylinders, respectively. Following Griffin and Ramberg \cite{griffin_vortex_street_1974}, the induced velocity in the horizontal direction can be estimated as:
\begin{equation}
U^*_{i,j}= -\frac{\Gamma_{TEV_{j}}}{l_j} \frac{\cosh2\pi\gamma_j(\sinh2\pi\gamma_j-\sin2\pi\epsilon_j \sinh2\pi\delta_j)}{sinh^2 2\pi\gamma_j sinh^22\pi\delta_j+\cos^22\pi\epsilon_j-2sinh2\pi\gamma_j \sin2\pi\epsilon_j \sinh2\pi\delta_j},
\label{eq_U^}
\end{equation}
where $\Gamma_{TEV_{j}}$ is the circulation of trailing edge vortex (TEV), $\gamma_j=v_j/2l_j$, $\epsilon=(x^*_{i,j}-c)/l_j$, and $\delta_j=y^*_{i,j}/l_j$. The terms $l_j$ and $v_j$ are the longitudinal and vertical spacings between the vortices in the vortex street of Foil $j$, respectively (see Fig.~\ref{fig_circulation}). They can be approximated as $l_j \approx U_{\infty}/f_j$ and $v_j \approx A_j.$ Hence, $U^*_{i,j}$ can be represented in a simpler form:
\begin{equation}
U^*_{i,j} \approx \sum\limits_{\substack{j=1 \\ j\neq i}}^n  \frac{\Gamma_{TEV_{j}} f_k}{U_{\infty}} \chi_{i,j}(f,A,U_{\infty}, x^*, y^*),
\label{eq_U^_simple}
\end{equation}
where $\chi_{i,j}$ is a vortex street function which depends on the oscillation frequency and amplitude, free-stream velocity, and longitudinal and vertical spacing between the foils. Note that  Griffin and Ramberg \cite{griffin_vortex_street_1974} also derived more complex formulation for the induced velocities by incorporating viscous decay of the vorticity term. However, we use inviscid solution since our derivation does not consider effects of $Re$. 

\subsection{Lift-based (circulatory) forces}
Now, we explain the derivation from Theodorsen's lift-based forces (circulatory) \cite{theodorsen-aeorodynamic-1935}. Oscillating foils produce unsteady lift as a result of its continuously varying angle-of-attack. Fig.~\ref{fig_velocity} illustrates velocity components generated and corresponding angles on a pitching foil in the two-foils system. The instantaneous angle-of-attack and the corresponding effective velocity of the foil is given by $\alpha_i=\theta_i - \arctan(V_{ind_{i}}/(U_{\infty}+U_{ind_{i}}))$ and $U_{eff_{i}}=\sqrt{V_{ind_{i}}^2+(U_{\infty}+U_{ind_{i}})^2}$, respectively. Lift-based forces in $x-$ direction, in $y-$ direction and moment about the leading edge are given as follows:
\begin{equation}
F_{x,L_{i}}=-L_{i} \sin(\theta_{i}-\alpha_{i})=-L_{i} V_{ind_{i}}/U_{eff_{i}},
\end{equation}
\begin{equation}
F_{y,L_{i}}=-L_{i} \sin(\theta_{i}-\alpha_{i})=-L_{i} (U_{\infty}+U_{ind_{i}})/U_{eff_{i}},
\end{equation}
\begin{equation}
M_{z,L_{i}}=-c L_{i}/4,
\end{equation}
where $L_{i} = (1/2)\rho U_{eff_{i}}^2 sc C_{L_{i}}$ and $C_{L_{i}} = 2 \pi \sin\alpha_{i} + (3/2)\pi \dot{\alpha_{i}}c/U_{\infty}$. 

Considering the induced velocities are very small compared to the free-stream velocity and the trailing edge velocity, we can approximate $\alpha_{i}$ as $\alpha_{i} \approx \theta_{i} -V_{ind_{i}}/(U_{\infty}+U_{ind_{i}}) \approx \theta_{i} -V_{ind_{i}}/U_{\infty}$. Consequently, $\dot{\alpha_{i}}= \dot{\theta_{i}}-\dot{V}_{ind_{i}}/U_{\infty}$. Now the lift-based forces and moment adopt the following forms: 

\begin{equation} 
F_{x,L_{i}} \sim \rho s c \left(\theta_{i} V_{ind_{i}} U_{eff} - \frac{V_{ind_{i}}^2}{U_{\infty}} U_{eff_{i}}+c \dot{\theta_{i}} V_{ind_{i}} - c \frac{\dot{V}_{ind_{i}}}{U_{\infty}} V_{ind_{i}}\right)
\end{equation}
\begin{equation} 
F_{y,L_{i}} \sim \rho s c \left(\theta U_{\infty} U_{eff_{i}}- V_{ind_{i}} U_{eff}+c \dot{\theta_{i}} U_{eff_{i}} - c \dot{V}_{ind_{i}}\right)
\end{equation}
\begin{equation} 
M_{z,L_{i}} \sim \rho s c^2 \left(\theta U_{eff_{i}}^2 - \frac{V_{ind_{i}}}{U_{\infty}} U_{eff_{i}}^2 +c \dot{\theta_{i}} \frac{U_{eff_{i}}^2}{U_{\infty}} - c \dot{V}_{ind_{i}} \frac{U_{eff_{i}}^2}{U_{\infty}^2}\right)
\end{equation}

\subsection{Added mass forces}
Next, we proceed with formulating the added mass forces, acting on the foils, by following Sedov's approach \cite{sedov-two-dimensional-1965}. Tangential force ($F_t$), normal force ($F_n$), and moment ($M_z$) on the foils are: 
\begin{equation} 
F_{t,AM_{i}} = \rho \pi s c \left(\frac{c}{4} V_{i} \dot{\theta}- \frac{c^2}{8} \dot{\theta_{i}}^2 \right)
\end{equation}
\begin{equation} 
F_{n,AM_{i}} = \rho \pi s c \left(\frac{c}{4} \dot{V_{i}} + \frac{c^2}{8} \ddot{\theta_{i}} \right)
\end{equation}
\begin{equation} 
F_{n,AM_{i}} = \rho \pi s c^2 \left(\frac{c}{8} \dot{V_{i}}-\frac{9c^2}{128} \ddot{\theta_{i}} \frac{1}{4}U_{i}V_{i}+\frac{c}{8}U_{i} \dot{\theta_{i}}\right)
\end{equation}
\noindent where $U_{i}$ and $V_{i}$ are normal and tangential velocity components impinging on Foil $i$, given as $U_{i}=V_{ind_{i}} \sin\theta_{i} + (U_{\infty}+U_{ind_{i}}) \cos\theta_{i}$ and $V_{i}=V_{ind_{i}} \cos\theta_{i} - (U_{\infty}+U_{ind_{i}}) \sin\theta_{i}$, respectively. In $x-y$ coordinates, the forces and the moment reduce to
\begin{dmath} 
F_{x,AM_{i}} \sim \rho s c \left[c \dot{\theta_{i}} V_{ind_{i}} (1+\theta^2_{i})+c \theta_{i} \dot{\theta_{i}} U_{\infty}+ c \theta_{i} \dot{\theta_{i}} U_{ind_{i}} + c^2 \dot{\theta_{i}}^2 + c \theta_{i} \dot{V}_{ind_{i}} +c \theta^2_{i} \dot{U}_{ind_{i}} +c^2 \theta_{i} \ddot{\theta_{i}}\right]
\end{dmath}
\begin{dmath} 
F_{y,AM_{i}} \sim \rho s c \left[c \theta_{i} \dot{\theta_{i}} V_{ind_{i}} +c \dot{\theta_{i}}U_{\infty}(1+\theta^2_{i}) + c \dot{\theta_{i}}U_{ind_{i}}(1+\theta^2_{i}) + c^2 \theta_{i} \dot{\theta_{i}}^2 + c \dot{V}_{ind_{i}}+ c \theta_{i} \dot{U}_{ind_{i}} +c^2 \ddot{\theta_{i}}\right]
\end{dmath}
\begin{dmath} 
M_{z,AM_{i}} \sim \rho s c^2 \left[c \dot{V}_{ind_{i}} +c \theta_{i} \dot{\theta_{i}} V_{ind_{i}} +c \dot{\theta_{i}} U_{\infty} +c \dot{\theta_{i}} U_{ind_{i}} +c^2 \ddot{\theta_{i}} + V_{ind_{i}} U_{\infty} (1+\theta^2_{i}) + V_{ind_{i}} U_{ind_{i}} (1+\theta^2_{i}) + \theta_{i} V_{ind_{i}}^2 + \theta_{i} U_{\infty}^2 + \theta_{i} U_{\infty} U_{ind_{i}}+ \theta U_{ind_{i}}^2\right]
\end{dmath}

\subsection{Derivation of cycle-averaged coefficients}
Now, we combine the lift-based and added mass forces and moments to determine the scaling relations for schooling foils. Following  Van Buren et al. \cite{buren-scaling-2019}, we approximate that $1+\theta^2 \approx 1$, considering $\theta$ is small and its contribution is negligible. The thrust and lift forces on the foils can be expressed in the following form:

\begin{dmath} 
F_{x_{i}} \sim \rho s c \left(\theta V_{ind_{i}} U_{eff_{i}} - \frac{V_{ind_{i}}^2}{U_{\infty}} U_{eff_{i}}+c \dot{\theta_{i}} V_{ind_{i}} - c \frac{\dot{V}_{ind_{i}}}{U_{\infty}} V_{ind_{i}}+c \theta_{i} \dot{\theta_{i}} U_{\infty}+ c \theta_{i} \dot{\theta_{i}} U_{ind_{i}} + c^2 \dot{\theta_{i}}^2 + c \theta_{i} \dot{V}_{ind_{i}} +c \theta^2_{i} \dot{U_{ind_{i}}} +c^2 \theta_{i} \ddot{\theta_{i}}\right) -F_{D_{i}}
\label{eq_thrust},
\end{dmath}

\begin{dmath} 
F_{y_{i}} \sim \rho s c \left(\theta_{i} U_{\infty} U_{eff_{i}}+ V_{ind_{i}} U_{eff_{i}}+c \dot{\theta_{i}} U_{eff_{i}} + c \dot{V}_{ind_{i}}+c \theta_{i} \dot{\theta_{i}} V_{ind_{i}} +c \dot{\theta_{i}}U_{\infty} + c \dot{\theta_{i}}U_{ind_{i}}+ c^2 \theta_{i} \dot{\theta_{i}}^2 + c \theta_{i} \dot{U}_{ind_{i}} +c^2 \ddot{\theta_{i}}\right)
\label{eq_lift},
\end{dmath}
\noindent where $F_D$ is the drag force on the foil. The drag experienced by a bluff body is proportional to the frontal area, $A_f$, which scales with $\sim U_{\infty}^2 A_f$. In the absence of large leading edge flow separation, drag on a pitching foil can be considered to be akin to that of a stationary foil. In this regard, frontal area of a stationary foil can be replaced with an averaged frontal area over a pitching cycle, which is proportional to the pitching amplitude, $A$ \cite{floryan_large_amplitude_2019, floryan-efficient-cruising-2018}. This relationship is supported by a quantitative evidence presented by Van Buren et al.\cite{buren-scaling-2019}, who demonstrated that the drag was a linear function of pitching amplitude and independent of heaving amplitude. Hence, drag offset should scale as $F_D \sim U_{\infty}^2 A$. Following this analogy, one can argue that the drag offset should also be influenced by $Re$ because drag on a bluff body varies with $Re$. In fact, Senturk and Smits\cite{senturk-reynolds-scaling-2019} already displayed that the drag term of pitching foils depended on the foil's thickness, oscillation amplitude, and $Re$, further supporting this analogy. They also provided a Reynolds number-based scaling for pitching foils by incorporating $Re$ into the drag-related term. In this study, on the other hand, we only consider the effect of the pitching amplitude, as the geometry of the foil is kept the same, and $Re$ remains fixed at $4000$.  This choice is based on our earlier work \cite{gungor-scaling}, where we revealed that the performance of pitching foils in side-by-side configurations were not significantly influenced by $Re$ when it exceeded $4000$. This observation is also biologically relevant considering that the natural swimming $Re$ of fish, reptiles, and cetaceans most of which typically fall within the range of $10^4-10^7$\cite{triantafyllou-optimal-thrust-1993, gazzola_aquatic_locomotion_2014}.

Next, we proceed with the derivation of the scaling relation for power consumed by tethered foils to continue producing propulsive forces. Power for simultaneously heaving and pitching foils is $P=F_y \dot{h}+M_z \dot{\theta}$ \cite{verma_vortex_evolution}. However, it can be approximated here as $P=F_y V_{ind}+M_z \dot{\theta}$, because heave velocity of the foil, $\dot{h}$, is akin to the induced velocity in the vertical direction, $V_{ind}$. 
\begin{dmath} 
P_{i} \sim \rho s c \left(\theta_{i} U_{\infty} U_{eff_{i}} V_{ind_{i}}+ V_{ind_{i}}^2 U_{eff_{i}}+c \dot{\theta_{i}} V_{ind_{i}} U_{eff_{i}} + c V_{ind_{i}} \dot{V}_{ind_{i}}+c \theta_{i} \dot{\theta_{i}} V_{ind_{i}}^2 +c \dot{\theta_{i}} V_{ind_{i}} U_{\infty} + c \dot{\theta_{i}} V_{ind_{i}} U_{ind_{i}}+ c^2 \theta_{i} \dot{\theta_{i}}^2 V_{ind_{i}} + c \theta_{i} V_{ind_{i}} \dot{U}_{ind_{i}} +c^2 \ddot{\theta_{i}} V_{ind_{i}} + c^2 \dot{\theta_{i}} \dot{V}_{ind_{i}}   +c^2 \dot{\theta_{i}}^2 U_{\infty} +c^2 \dot{\theta_{i}}^2 U_{ind_{i}} +c^3 \dot{\theta_{i}}\ddot{\theta_{i}}  + c\dot{\theta_{i}} V_{ind_{i}} U_{ind_{i}} + c\theta_{i} \dot{\theta_{i}} U_{\infty}^2 + c\theta_{i} \dot{\theta_{i}} U_{\infty} U_{ind_{i}}+ c\theta_{i} \dot{\theta_{i}} U_{ind_{i}}^2 \right)
\label{eq_power}.
\end{dmath}

\begin{figure}
	\centering
	\includegraphics[width=1\textwidth]{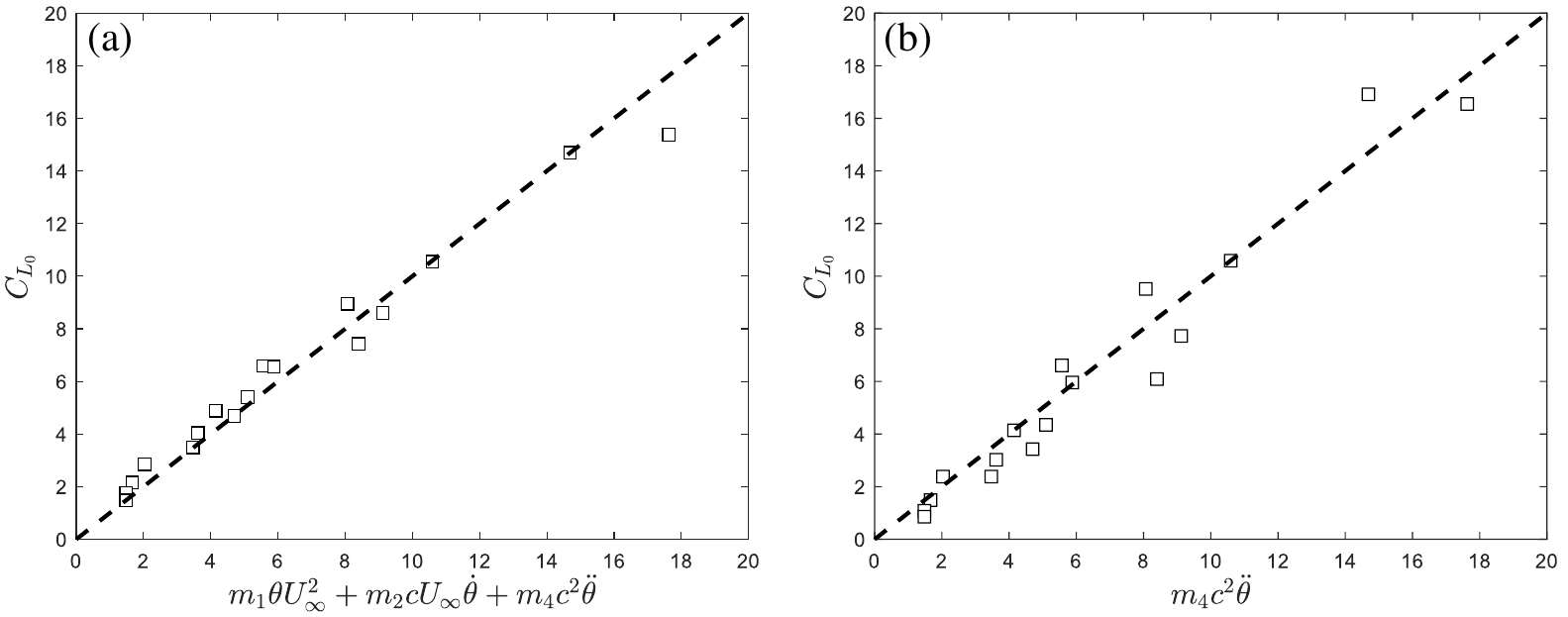}
	\caption{Coefficient of unsteady lift amplitude, $C_{L_{0}}=L_0/\frac{1}{2} \rho U_{\infty}^2 s c$, scaled with (a) $m_1 \theta U_{\infty}^2 + m_2 c U_{\infty} \dot{\theta} +m_4 c^2 \ddot{\theta}$ and (b) $m_4 c^2 \ddot{\theta}$ where $m_{1-4}$ are empirical coefficients determined through a least-squares linear regression analysis over randomly selected data points.}
	\label{fig_lift_empirical}
\end{figure}

To finalize these laws, we should determine how $V_{ind}$ and $U_{ind}$ scale with the flow quantities. As previously explained, induced velocities arise from the circulation of the foils, which is related to the generated lift (see Eq.~\ref{eq_v_ind} and Eq.~\ref{eq_u_ind}). Therefore, we approximate $L_0$ from the equation for the lift force (Eq.~\ref{eq_lift}). To simplify the approximation, only a pure pitching motion is considered, reducing Eq.~\ref{eq_lift} to $F_{y_{i}} \sim \theta_{i} U_{\infty}^2 +c \dot{\theta_{i}}U_{\infty} + c^2 \theta_{i} \dot{\theta_{i}}^2 + c^2 \ddot{\theta_{i}}$. Upon expansion, it is found that the third term can be neglected by employing the small angle approximation since it contains $\theta_0^3$, whereas the other terms contain $\theta_0$. The equation is further simplified by assuming that the fourth term, which scales as $c^2 \ddot{\theta}$, is the dominant term. To confirm this assumption,  we analyzed the \ah{dominance} of $c^2 \ddot{\theta}$ term to represent unsteady lift amplitude data alone. As illustrated in Fig.~\ref{fig_lift_empirical}, $c^2 \ddot{\theta}$ provides an adequate approximation for the amplitude of unsteady lift, even though inclusion of the terms $\theta U_{\infty}^2$ and $c \dot{\theta}U_{\infty}$ improves the fit. Thus, we assume that the amplitude of lift would scale as $L_0 \sim f^2 c^2 \theta_0$. Similarly, we need an approximation for $\Gamma_{TEV}$, which was presented by Schnipper et al.\cite{schnipper-vortex-wakes-2009} as:
\begin{equation}
\Gamma_{TEV_{i}}=\frac{1}{2} \int_{0}^{1/2f} V^2_{TE_{i}} dt \approx \frac{1}{2}\pi^2 f c^2 \theta^2_0,	
\label{eq_TEV}
\end{equation}
where $V_{TE}$ is the trailing edge velocity magnitude, calculated as $V_{TE_{i}}=c \dot{\theta_{i}}$. This elucidates that $\Gamma_{TEV}$ scales as $\Gamma_{TEV} \sim f c^2 \theta^2_0$ \cite{buren_flow_speed_2018}. Next, we integrate the thrust and power relations over a pitching cycle and non-dimensionalize them through Eq.~\ref{eq_ct} and Eq.~\ref{eq_cp} to obtain $\overline{C_{T}}$ and $\overline{C_{P}}$, respectively. We assume cycle-averaged effective velocity to be similar to the free-stream flow, \textit{i.e.,} $\overline{U_{eff}} \approx U_{\infty}$ as its contribution would be negligible compared to other flow quantities\cite{buren-bio-inspired-2019}. Hence,  $\overline{CT_i}$ and $\overline{CP_i}$ can be expressed as:
\begin{dmath}
 \overline{C_{T,i}}=ct_1 \Lambda_{T1_{i}}+ct_2 \Lambda_{T2_{i}}+ct_3 \Lambda_{T3_{i}}+ct_4 \Lambda_{T4_{i}}+ct_5 \Lambda_{T5_{i}}\\+ct_6 \Lambda_{T6_{i}}+ct_7 \Lambda_{T7_{i}}+ct_8 \Lambda_{T8_{i}}+ct_9 \Lambda_{T9_{i}}+ct_{10} \Lambda_{T10_{i}},
 \label{eq_ct_i}
\end{dmath}
\begin{dmath}
 \overline{C_{P,i}}=cp_1 \Lambda_{P1_{i}}+cp_2 \Lambda_{P2_{i}}+cp_3 \Lambda_{P3_{i}}+cp_4 \Lambda_{P4_{i}}+cp_5 \Lambda_{P5_{i}}+cp_6 \Lambda_{P6_{i}}+cp_7 \Lambda_{P7_{i}}+cp_8 \Lambda_{P8_{i}}+cp_9 \Lambda_{P9_{i}}+cp_{10} \Lambda_{P10_{i}}+cp_{11} \Lambda_{P11_{i}}+cp_{12} \Lambda_{P12_{i}}+cp_{13} \Lambda_{P13_{i}}+cp_{14} \Lambda_{P14_{i}}+cp_{15} \Lambda_{P15_{i}}+cp_{16} \Lambda_{P16_{i}}+cp_{17} \Lambda_{P17_{i}}+cp_{18} \Lambda_{P18_{i}}+cp_{19} \Lambda_{P19_{i}}.
  \label{eq_cp_i}
\end{dmath}
where definitions of expressions $\Lambda_{T1-10}$ and $\Lambda_{P1-19}$ are given in Table \ref{table_thrust} and Table \ref{table_power}, respectively.

\begin{table*}
	\caption{\label{table_thrust} Definitions of the thrust terms in the derived scaling equations.}
	\centering
	\renewcommand{\arraystretch}{1.4}
	\setlength{\tabcolsep}{8.2pt}
	\begin{tabular}{ll}
            \hline
            Thrust Term &  Definition \\
		\hline
$\Lambda_{T1_{i}}$ & $\sum\limits_{\substack{j=1 \\ j\neq i}}^n St^2 \cos(\xi_j+\phi_j-\phi_i)\left(\frac{\cos\psi_{f_{i,j}}}{r_{f_{i,j}}}-\frac{\cos\psi_{w_{i,j}}}{r_{w_{i,j}}} \right)$ \\
$\Lambda_{T2_{i}}$ & $\sum\limits_{\substack{j=1 \\ j\neq i}}^n St^2f^{*2} \left(\frac{\cos\psi_{f_{i,j}}}{r_{f_{i,j}}}-\frac{\cos\psi_{w_{i,j}}}{r_{w_{i,j}}} \right)^2$\\
$\Lambda_{T3_{i}}$ & $\sum\limits_{\substack{j=1 \\ j\neq i}}^n St^2 f^* \sin (\xi_j+\phi_j-\phi_i) \left(\frac{\cos\psi_{f_{i,j}}}{r_{f_{i,j}}}-\frac{\cos\psi_{w_{i,j}}}{r_{w_{i,j}}} \right)$\\
$\Lambda_{T4_{i}}$ & $\sum\limits_{\substack{j=1 \\ j\neq i}}^n \underline {St f^{*4} \left(\frac{\cos\psi_{f_{i,j}}}{r_{f_{i,j}}}-\frac{\cos\psi_{w_{i,j}}}{r_{w_{i,j}}}\right)^2}$\\
$\Lambda_{T5_{i}}$ & $St^2$\\
$\Lambda_{T6_{i}}$ & $\sum\limits_{\substack{j=1 \\ j\neq i}}^n St^2 f^* \sin (\xi_j+\phi_j-\phi_i) \left(\frac{\sin\psi_{f_{i,j}}}{r_{f_{i,j}}}-\frac{\sin\psi_{w_{i,j}}}{r_{w_{i,j}}}\right)$\\
$\Lambda_{T7_{i}}$ & $\underline{St A^*}$\\
$\Lambda_{T8_{i}}$ & $\sum\limits_{\substack{j=1 \\ j\neq i}}^n \underline{St^3 \left(\frac{\sin\psi_{f_{i,j}}}{r_{f_{i,j}}}-\frac{\sin\psi_{w_{i,j}}}{r_{w_{i,j}}}\right)}$\\
$\Lambda_{T9_{i}}$ & $\sum\limits_{\substack{j=1 \\ j\neq i}}^n \underline{St^3 A^{*}\chi_{i,j}}$\\
$\Lambda_{T10_{i}}$  & $A^*$\\	
		\hline
	\end{tabular}
\end{table*}

\begin{table*}
	\caption{\label{table_power} Definitions of the power terms in the derived scaling equations.}
	\centering
	\renewcommand{\arraystretch}{1.4}
	\setlength{\tabcolsep}{8.2pt}
	\begin{tabular}{ll}
            \hline
            Power Term &  Definition \\
		\hline
$\Lambda_{P1_{i}}$ &$ \sum\limits_{\substack{j=1 \\ j\neq i}}^n St^2 \cos(\xi_j+\phi_j-\phi_i) \left(\frac{\cos\psi_{f_{i,j}}}{r_{f_{i,j}}}-\frac{\cos\psi_{w_{i,j}}}{r_{w_{i,j}}}\right)$\\
$\Lambda_{P2_{i}} $&$ \sum\limits_{\substack{j=1 \\ j\neq i}}^n St^4 \left(\frac{\cos\psi_{f_{i,j}}}{r_{f_{i,j}}}-\frac{\cos\psi_{w_{i,j}}}{r_{w_{i,j}}}\right)^2$\\
$\Lambda_{P3_{i}} $&$ \sum\limits_{\substack{j=1 \\ j\neq i}}^n St^2 f^{*} \sin(\xi_j+\phi_j-\phi_i)\left(\frac{\cos\psi_{f_{i,j}}}{r_{f_{i,j}}}-\frac{\cos\psi_{w_{i,j}}}{r_{w_{i,j}}}\right)$\\
$\Lambda_{P4_{i}} $&$ \sum\limits_{\substack{j=1 \\ j\neq i}}^n \underline{St^2 f^{*3} \left(\frac{\cos\psi_{f_{i,j}}}{r_{f_{i,j}}}-\frac{\cos\psi_{w_{i,j}}}{r_{w_{i,j}}}\right)^2}$\\
$\Lambda_{P5_{i}} $& $\sum\limits_{\substack{j=1 \\ j\neq i}}^n St^4 f^{*} \sin(2\xi_j+2\phi_j-2\phi_i)\left(\frac{\cos\psi_{f_{i,j}}}{r_{f_{i,j}}}-\frac{\cos\psi_{w_{i,j}}}{r_{w_{i,j}}}\right)^2$\\
$\Lambda_{P6_{i}} $&$ \sum\limits_{\substack{j=1 \\ j\neq i}}^n \underline{St^3 f^{*2} \left(\frac{\cos\psi_{f_{i,j}}}{r_{f_{i,j}}}-\frac{\cos\psi_{w_{i,j}}}{r_{w_{i,j}}}\right)\left(\frac{\sin\psi_{f_{i,j}}}{r_{f_{i,j}}}-\frac{\sin\psi_{w_{i,j}}}{r_{w_{i,j}}}\right)}$\\
$\Lambda_{P7_{i}} $&$ \sum\limits_{\substack{j=1 \\ j\neq i}}^n St^4 \cos(\xi_j+\phi_j-\phi_i)\left(\frac{\cos\psi_{f_{i,j}}}{r_{f_{i,j}}}-\frac{\cos\psi_{w_{i,j}}}{r_{w_{i,j}}}\right)$\\
$\Lambda_{P8_{i}} $& $\sum\limits_{\substack{j=1 \\ j\neq i}}^n St^2 f^{*2} \cos(\xi_j+\phi_j-\phi_i)\left(\frac{\cos\psi_{f_{i,j}}}{r_{f_{i,j}}}-\frac{\cos(\psi_{w_{i,j}})}{r_{w_{i,j}}}\right)$\\
$\Lambda_{P9_{i}} $&$ St^2$\\
$\Lambda_{P10_{i}} $&$ \sum\limits_{\substack{j=1 \\ j\neq i}}^n \underline{St^3 f^{*} \left(\frac{\sin\psi_{f_{i,j}}}{r_{f_{i,j}}}-\frac{\sin\psi_{w_{i,j}}}{r_{w_{i,j}}}\right)}$\\
$\Lambda_{P11_{i}}$ &$ \underline{St^2 f^{*}}$\\
$\Lambda_{P12_{i}} $& $\underline{St A^*}$\\
$\Lambda_{P13_{i}} $&$ \sum\limits_{\substack{j=1 \\ j\neq i}}^n \underline{St^3 \left(\frac{\sin\psi_{f_{i,j}}}{r_{f_{i,j}}}-\frac{\sin\psi_{w_{i,j}}}{r_{w_{i,j}}}\right)}$\\
$\Lambda_{P14_{i}} $& $\sum\limits_{\substack{j=1 \\ j\neq i}}^n St^4 f^{*} \cos(2\xi_j+2\phi_j-2\phi_i)\left(\frac{\sin\psi_{f_{i,j}}}{r_{f_{i,j}}}-\frac{\sin\psi_{w_{i,j}}}{r_{w_{i,j}}}\right)^2$\\
$\Lambda_{P15_{i}} $&$ \sum\limits_{\substack{j=1 \\ j\neq i}}^n St^4 f^* \sin(\xi_j+\phi_j-\phi_i)\left(\frac{\cos\psi_{f_{i,j}}}{r_{f_{i,j}}}-\frac{\cos\psi_{w_{i,j}}}{r_{w_{i,j}}}\right)\chi_{i,j}$\\
$\Lambda_{P16_{i}} $& $\sum\limits_{\substack{j=1 \\ j\neq i}}^n St^4 \chi_{i,j}$\\
$\Lambda_{P17_{i}} $&$ \sum\limits_{\substack{j=1 \\ j\neq i}}^n \underline{St^3 A^* \chi_{i,j}}$\\
$\Lambda_{P18_{i}} $& $\sum\limits_{\substack{j=1 \\ j\neq i}}^n \underline{St^5 \left(\frac{\sin\psi_{f_{i,j}}}{r_{f_{i,j}}}-\frac{\sin\psi_{w_{i,j}}} {r_{w_{i,j}}}\right)\chi_{i,j}}$\\
$\Lambda_{P19_{i}} $&$ \sum\limits_{\substack{j=1 \\ j\neq i}}^n \underline{St^5 A^*\chi^2_{i,j}}$\\	
		\hline
	\end{tabular}
\end{table*}

\section{Results}
\label{results}

We now continue with computing the terms in Eq.~\ref{eq_ct_i} and Eq.~\ref{eq_cp_i} to assess the effectiveness of the derived equations in capturing the fundamental flow physics. For the purpose of this analysis, we focus on the case of two-foils systems, since these configurations represent the majority of our dataset. To calculate the equations for the two foils, we simply substitute $i=1$ and $i=2$ to represent Foils 1 and 2, respectively, while keeping the value of n fixed at 2. \rev{Note that $j$ can get only one value since it represents the contribution due to the other foil in the system, \textit{i.e.,} $j=2$ for $i=1$ and $j=1$ for $i=2$.} Here, $ct_{1-10}$ and $cp_{1-19}$ are scaling coefficients that are determined numerically. In accordance with Floryan et al. \cite{floryan-scaling-propulsive-performance-2017} and Van Buren et al. \cite{buren-scaling-2019}, we underline inherently out-of-phase terms, such as displacement-velocity or velocity-acceleration terms, \textit{i.e.,} $\theta-\dot{\theta}$ or $\dot{\theta}-\ddot{\theta}$. These terms have $90^\circ$ phase lag between them; therefore, a time-averaged quantity of their product should be very small. However, it is possible for strong non-linear effects to exist in such complex flows as suggested by Liu et al. \cite{liu_unsteady_airfoil_2015}, which may break the inherent out-of-phase synchronization of these terms. Hence, we chose to keep these terms in our analysis. 

\subsection{Two-Foils Systems}

 \begin{figure}
	\centering
	\includegraphics[width=1\textwidth]{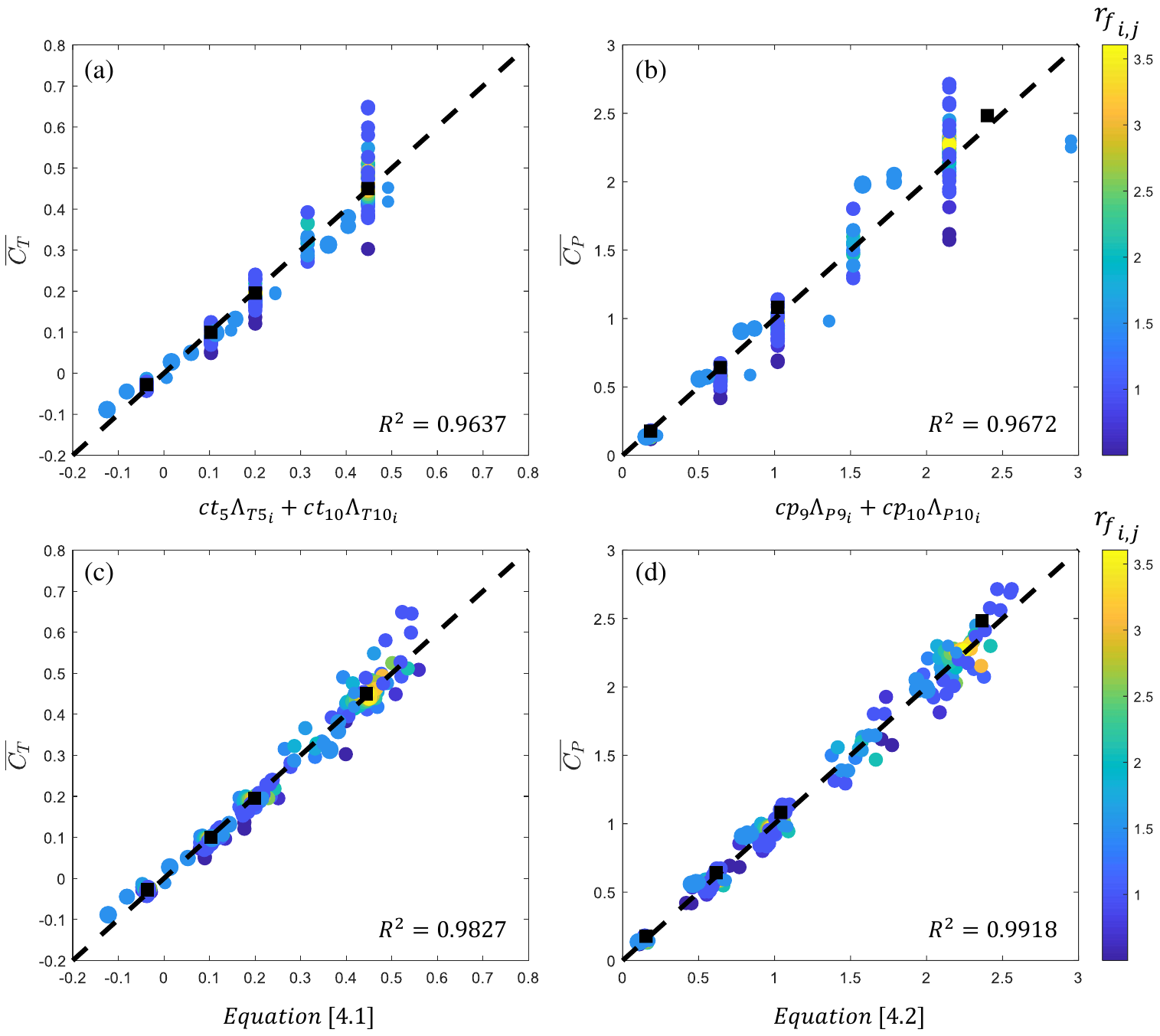}
	\caption{(a-c) Thrust and (b-d) power scaling results of pitching foils in two-foil configurations for varying pitching amplitude and phase difference at $Re=4000$. Numerical data is plotted against (a-b) only pure-pitching terms, (c-d) full scaling equations. Color and size indicates the radial distance between the foils ($|r_{f_{i,j}}|$) and pitching amplitude ($A^*$), respectively. Isolated foil data is illustrated with black squares. }
	\label{fig_scaling}
\end{figure}

In order to determine the scaling coefficients, $ct_{1-10}$ and $cp_{1-19}$, least-squares linear regression analysis is employed over the entire dataset. Ultimately, it is noticed that some terms could be eliminated from the formulations as their impact on the results is marginal for the given parameter space. This facilitates the establishment of more simplified scaling equations with little compromise on accuracy. Consequently, the terms associated with $ct_1$, $ct_4$, $ct_6$, $ct_7$, $ct_8$, and $ct_9$ in the equations for thrust, and $cp_1$, $cp_4$, $cp_5$, $cp_{12}$, $cp_{13}$, $cp_{14}$, $cp_{15}$, $cp_{16}$, $cp_{17}$, and $cp_{19}$ in the equations for power are excluded from the final form of the scaling equations:
\begin{dmath}
\overline{C_{T,i}}=\underbrace {ct_2 \Lambda_{T2_{i}}+ct_3 \Lambda_{T3_{i}}}_{\text {induced velocity}}+ \underbrace{ct_5 \Lambda_{T5_{i}}+ct_{10} \Lambda_{T10_{i}}}_{\text {pure-pitching}}
 \label{eq_scaling_ct},
\end{dmath}
\begin{dmath}
 \overline{C_{P,i}}=\underbrace{cp_2 \Lambda_{P2_{i}}+cp_3 \Lambda_{P3_{i}}+cp_6 \Lambda_{P6_{i}}+cp_7 \Lambda_{P7_{i}}+cp_8 \Lambda_{P8_{i}}+cp_{10} \Lambda_{P10_{i}}+cp_{18} \Lambda_{P18_{i}}}_{\text{induced velocity}} + \underbrace{cp_{9} \Lambda_{P9_{i}}+cp_{11} \Lambda_{P11_{i}}}_{\text{pure-pitching}}.
 \label{eq_scaling_cp}
\end{dmath}
 The numerical coefficients are found to be $ct_2=0.30$, $ct_3=-0.84$, $ct_5=3.49$, and $ct_{10}=-0.41$ for thrust scaling and $cp_2=5.65$, $ct_3=-1.15$, $ct_6=-2.03$, $cp_7=135.76$, $cp_8=-11.95$, $cp_9=1.57$, $cp_{10}=2.23$, $cp_{11}=8.76$, and $cp_{18}=3.96$. 

Terms in the scaling equations (Eq.~\ref{eq_scaling_ct} and Eq.~\ref{eq_scaling_cp}) are classified based on their origin. The "pure-pitching" terms arise from the pitching oscillations of the foils, while the terms for "induced velocity" stem from the interactions between the foils. The pure-pitching terms are plotted alone against the numerical simulation data for the coefficients of thrust and power in Fig.~\ref{fig_scaling}a and Fig.~\ref{fig_scaling}b, respectively. The figures also comprise the performance of isolated foils, which are essentially for the foils uninfluenced by the effects of induced velocity. They fail to achieve a satisfactory collapse, except for the isolated foils and two-foils systems with significant separation distance. This indicates the insufficiency of the pure-pitching terms in accurately representing complex flows with intense foil-foil interactions. On the other hand, Figs.~\ref{fig_scaling}c and ~\ref{fig_scaling}d display the fit of the data, where the numerical results demonstrate an excellent linear collapse on the complete scaling equations (Eq.~\ref{eq_scaling_ct} and Eq.~\ref{eq_scaling_cp}). It provides a compelling evidence that the derived terms proficiently capture the underlying flow physics. The improved r-squared values, depicting a notable shift from $R^2=0.9637$ to $R^2=0.9827$ for the thrust scaling and from $R^2=0.9672$ to $R^2=0.9918$ for the power scaling, unequivocally quantify the critical role played by the induced velocity terms. 

\begin{figure}
	\centering
	\includegraphics[width=0.5\textwidth]{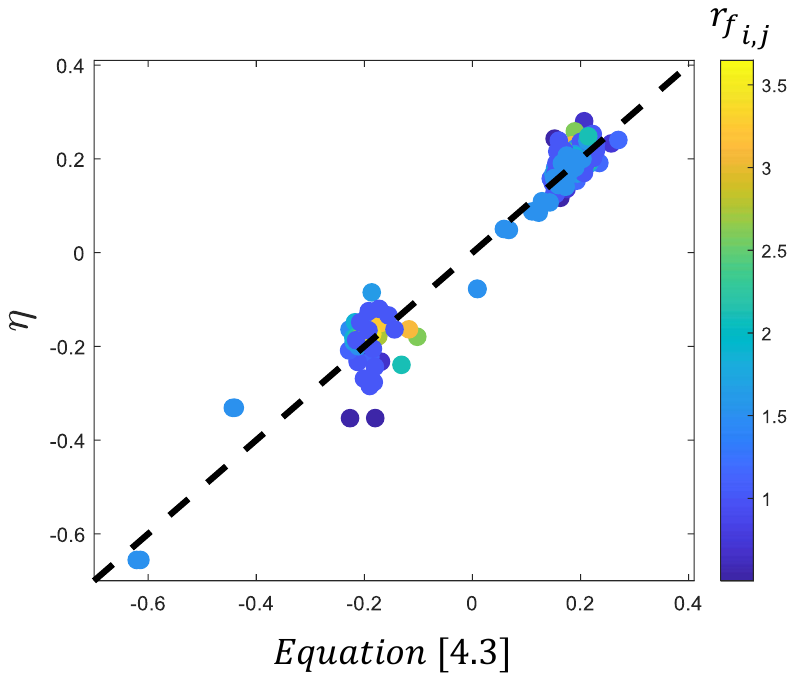}
	\caption{Simulation efficiency plotted against scaling efficiency for two-foil configurations at $Re=4000$. Color and size indicates the radial distance between the foils ($|r_{f_{i,j}}|$) and pitching amplitude ($A^*$), respectively. }
	\label{fig_scaling_efficiency}
\end{figure}

The results display certain outliers in Figs.~\ref{fig_scaling}c and ~\ref{fig_scaling}d, primarily noticeable in scenarios where the foils are in \ah{very close proximity. We categorize these configurations as interfering configurations, such that the leader and follower foils can no longer be considered separate bodies.} This observation is consistent with our expectations, given that the interaction between the foils exerts a substantial influence on their lift characteristics. In a previous study\cite{gungor_wake_merging}, we demonstrated that the formation of leading edge vortices is suppressed on the adjacent surfaces of the foils when placed in a side-by-side arrangement. This phenomenon directly influences the circulation generated by the foils and their wake, resulting in deviation from our initial assumption that the lift follows an ideal sine-wave, \textit{i.e.,} $L_i=L_0  \sin(2 \pi f t-\xi_i)$. We conjecture that the inclusion of a lift-correction term for close foil proximity could potentially  offer a solution. However, \ah{since the deviations are already relatively small, we do not consider changing the form of lift variation.}

 We calculate the propulsive efficiency of the foils utilizing the scaling equations, given as:
 \begin{equation}
    \eta_i=\frac{\overline{C_{T_{i}}}}{\overline{C_{P_{i}}}}=\frac{\mbox{RHS~of}~Eq.~\ref{eq_scaling_ct}}{\mbox{RHS~of}~Eq.~\ref{eq_scaling_cp}}
 \label{eq_scaling_eff}.
\end{equation}
However, when plotting efficiency from the scaling relations (Eq.~\ref{eq_scaling_eff}) against that from the computational data (Eq.~\ref{eq_eff}, as shown in Fig.~\ref{fig_scaling_efficiency}), we observe \ah{that the data has not collapse as well as those of the coefficients.} This outcome \ah{motivated} us to hypothesize that the traditional definition of efficiency (Eq.~\ref{eq_eff}) may be inadequate for schooling configurations. Therefore, it is necessary to redefine efficiency specifically for these setups. Similar to the previously proposed lift-correction term, the introduction of a moment-correction term could potentially resolve this issue. However, examining and including a moment-correction term falls outside the scope of this study, and we consider it as a potential avenue for our future research.

\subsection{Three-Foils Systems}

\begin{figure}
 \centering
	\includegraphics[width=0.35\textwidth]{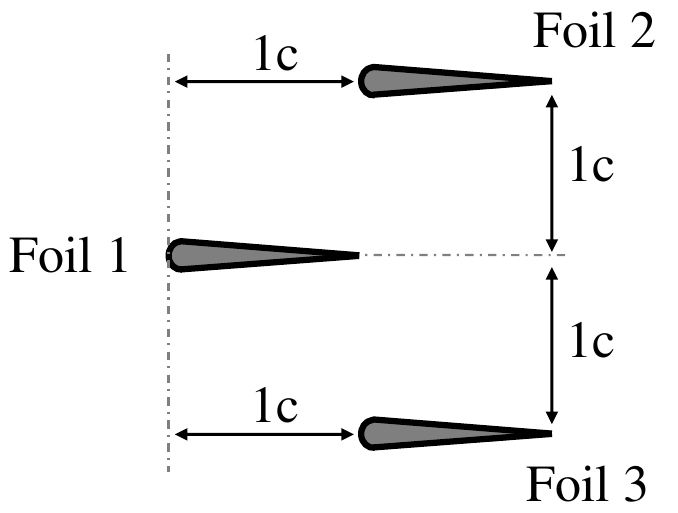}
	\caption{\rev{Schematics of the three-foil configuration (not to scale).} }
	\label{fig_3foil_configuration}
\end{figure}

\begin{figure}
	\centering
	\includegraphics[width=1\textwidth]{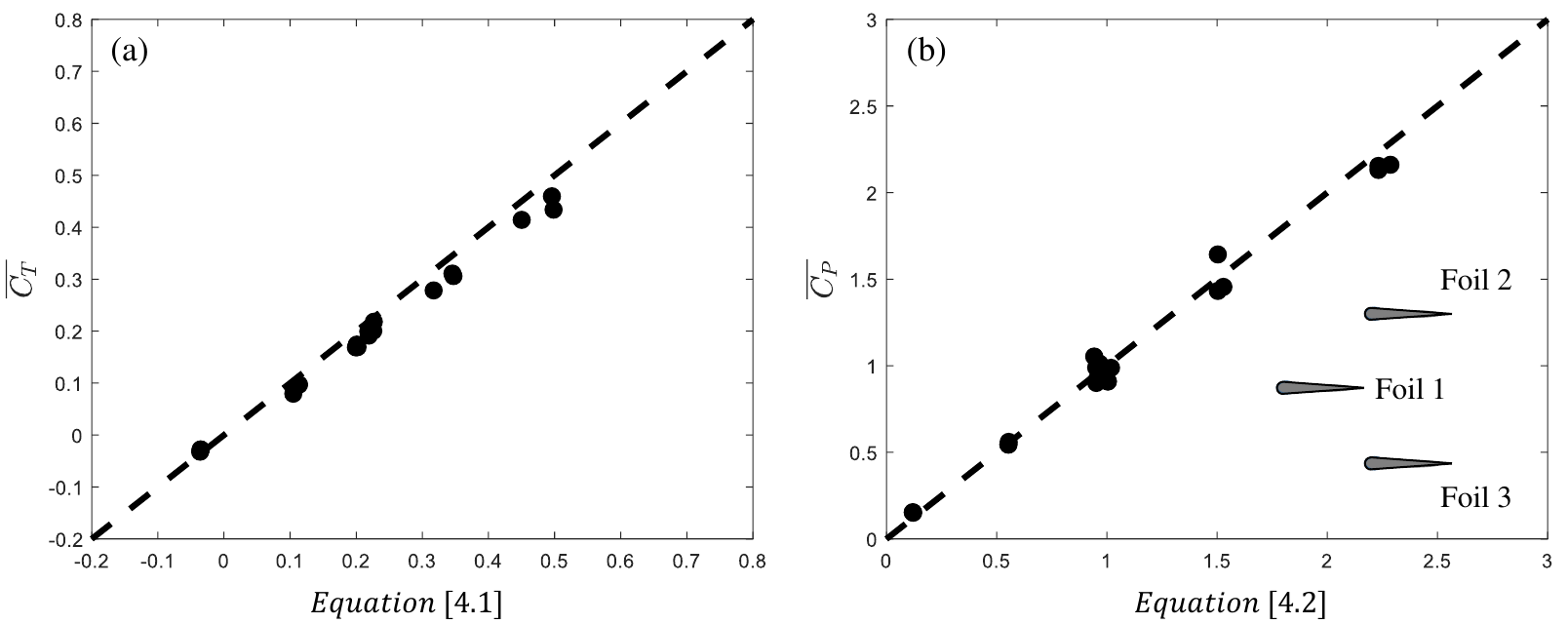}
	\caption{(a) Thrust and (b) power scaling results of pitching foils in a three-foil configuration at $Re=4000$. }
	\label{fig_scaling_3foil}
\end{figure}

While the scaling relations presented in this paper have primarily focused on two-foils configurations, we assert that they can be effectively extended to multi-foil systems as well. This stems from the accurate modelling of foil-foil interactions in accordance with the underlying flow physics. To validate this hypothesis, we conducted simulations involving a three-foil configuration. In this setup, two follower foils are positioned at a separation distance of $1c$ from the leader foil in both the horizontal ($x-$) and vertical ($y-$) directions \rev{as illustrated in Fig. \ref{fig_3foil_configuration}}. This specific configuration is inspired by the experiments of Ashraf et al. \cite{ashraf-syncronization-2016}, who observed it to represent the most probable schooling arrangement among other schools of three fish. The results of our simulations, as shown in Fig.~\ref{fig_scaling_3foil}, demonstrate an excellent alignment with the scaling equations for both thrust and power, despite employing the coefficients derived from the two-foil systems dataset without any modifications. This highlights the versatility and robustness of the derived scaling laws to estimate the propulsive performance of any multi-foil schooling configuration.

\section{Conclusions}
\label{conclusion}
The current study presents a novel approach \ah{in developing physics-informed} scaling relations that accurately estimate the propulsive performance metrics of pitching foils in various schooling formations. Building on the existing scaling laws established for single oscillating foils, this research extends their applicability to multi-foil systems by incorporating the intricate dynamics of foil-foil interactions. Our analysis considers vortex-induced velocities, imposed by a foil through its circulation on the other foils in the system. The vertical component of the induced velocity is analogous to the heaving motion, suggesting that pitching foils in multi-foil systems can be treated as simultaneously heaving and pitching bodies. The scaling relations unveil a novel, physics-based methodology for two or more oscillating foils, exhibiting outstanding agreement with numerical simulations. These relations account for the influence of the positioning of foils, phase differences between them, and the amplitudes and frequencies of oscillations. By providing a deeper understanding of the fundamental flow physics, the derived relations are expected to play a crucial role in offering vital insights for the design and optimization of more effective underwater propulsive and energy harvesting systems.

\vskip6pt








\bibliographystyle{unsrt}

\bibliography{references}



\end{document}